\documentclass[aps,prl,nofootinbib,amsfonts,tightenlines,twocolumn,superscriptaddress]{revtex4-1}
\makeatletter
\renewcommand{\p@subsection}{}
\renewcommand{\p@subsubsection}{}
\makeatother

\usepackage[T1]{fontenc}
\usepackage[utf8]{inputenc}
\usepackage{amsmath}
\usepackage{amsfonts}
\usepackage{amssymb}
\usepackage{amsthm}
\usepackage{caption}
\usepackage{multirow}
\usepackage[english]{babel}
\usepackage{fontenc}
\usepackage{graphicx}
 \usepackage{hyperref}
 \usepackage{dsfont}
  \usepackage{comment}

\usepackage{mathptmx}
\usepackage{float}
\usepackage[usenames,dvipsnames]{color}
\definecolor{mightnightblue}{RGB}{25,25,112}
\usepackage{hyperref}
 \hypersetup{
     colorlinks=true,
     linkcolor= Black,
     citecolor=mightnightblue,
     urlcolor=mightnightblue
     } 
\usepackage{caption}
\newcommand {\ignore}[1]{}

\newcommand{\sm}{{standard model }}

\def\cpv{CP violation }
\newcommand{\AddrAHEP}{AHEP Group, Institut de F\'{i}sica Corpuscular --
  C.S.I.C./Universitat de Val\`{e}ncia, Parc Cientific de Paterna.\\
  C/Catedratico Jos\'e Beltr\'an, 2 E-46980 Paterna (Val\`{e}ncia) - SPAIN}

\begin{document}

\preprint{IP/BBSR/2017-2}
\title{Cornering the revamped BMV model with neutrino oscillation data}
\date{\today}

\author{Sabya Sachi Chatterjee}\email{sabya@iopb.res.in}
\affiliation{Institute of Physics, Sachivalaya Marg, Sainik  School Post, Bhubaneswar 751005, India}
\affiliation{Homi Bhabha National Institute, Training School Complex, Anushakti Nagar, Mumbai 400085, India}
\author{Mehedi Masud}\email{masud@ific.uv.es}
\affiliation{\AddrAHEP}
\author{Pedro Pasquini}\email{pasquini@ifi.unicamp.br}
\affiliation{Instituto de F\'isica Gleb Wataghin - UNICAMP, {13083-859}, Campinas SP, Brazil}
\author{J.W.F. Valle}\email{valle@ific.uv.es, URL:  http://astroparticles.es/}
\affiliation{\AddrAHEP}

\begin{abstract}

  Using the latest global determination of neutrino oscillation
  parameters from~\cite{deSalas:2017kay} we examine the status of the
  simplest revamped version of the BMV (Babu-Ma-Valle) model, proposed
  in~\cite{Morisi:2013qna}.
  The model predicts a striking correlation between the ``poorly
  determined'' atmospheric angle $\theta_{23}$ and CP phase
  $\delta_{CP}$, leading to either maximal \cpv or none, depending on
  the preferred $\theta_{23}$ octants.
  We determine the allowed BMV parameter regions and compare with
  the general three-neutrino oscillation scenario.
  We show that in the BMV model the higher octant is possible only at
  99\% C.L., a stronger rejection than found in the general case.
  By performing quantitative simulations of forthcoming DUNE and T2HK
  experiments, using only the four ``well-measured'' oscillation
  parameters and the indication for normal mass ordering, we also map
  out the potential of these experiments to corner the model.
  The resulting global sensitivities are given in a robust form,
  that holds irrespective of the true values of the oscillation
  parameters. 

\end{abstract}
 \maketitle

 \section{Introduction}\label{sec:intro}
 
 The observed flavor structure of quarks and leptons is unlikely to be
 an accident.
 Specially puzzling are the neutrino oscillation
 parameters~\cite{deSalas:2017kay}, featuring two large angles with no
 counterpart in the quark sector~\cite{1674-1137-40-10-100001}, as
 well as a smaller mixing parameter measured at reactors, and which
 lies suspiciously close in magnitude to the Cabbibo
 angle~\cite{Boucenna:2012xb,Roy:2014nua}.
While the \sm gives an incredibly good description of ``vertical'' or
intrafamily gauge interactions, it gives no guidance concerning
``horizontal'' interfamily interactions.
A reasonable attempt to shed light on the pattern of fermion masses
and mixings is the idea of flavor
symmetry~\cite{Hirsch:2012ym,Morisi:2012fg,ishimori2012introduction}.
Over the last years many models have been proposed in order to account
for the pattern of neutrino
oscillations~\cite{Morisi:2012fg,King:2014nza} and most of them make well-defined predictions for the ``poorly
determined'' oscillation parameters $\sin^2\theta_{23}$ and
$\delta_{CP}$~\cite{Chen:2015jta,Pasquini:2016kwk,CentellesChulia:2017koy,CarcamoHernandez:2017owh}.

In this paper we consider, for definiteness, on the model suggested
in~\cite{Morisi:2013qna}, i.e. the simplest flavon generalization of
the  $A_4$-symmetry-based BMV model~\cite{Babu:2002dz}.
This revamped model predicts a sharp correlation between the CP phase
and the atmospheric angle $\theta_{23}$, which implies either maximal
\cpv or none, depending on the preferred octants of the atmospheric
angle $\theta_{23}$.
We focus on the capability of future experiments
DUNE~\cite{Acciarri:2015uup} and T2HK~\cite{Abe:2015zbg} to test the
predictions of the simplest realistic $A_4$ model presented
in~\cite{Morisi:2013qna} given the current measurements of the
oscillation parameters.
We also perform quantitative simulations of the future DUNE and T2HK
experiments in order to illustrate their potential in testing the
model.
To this endeavor we use only the four ``well-measured'' oscillation
parameters plus the indication in favor of normal mass ordering
and lower octant.
We determine their increased sensitivity in probing the BMV model
compared to the general unconstrained case.
We present the results as robust, global model-testing criteria that
hold for any choice of the true values of the oscillation parameters.

\section{Theoretical preliminaries}
\label{sec:theor-prel}

The model is a minimal extension of the BMV model~\cite{Babu:2002dz},
which assembles the $SU(2)_L$ doublet fermions into an $A_4$ triplet
within a supersymmetric framework. It requires the existence of extra
heavy fermions and three scalars $\chi_i$, $i=1,2,3$, all of them
belonging to $A_4$ triplets representation and coupled through
standard Yukawa interactions.
Both standard Higgs fields $H_i$ and the three new scalars $\chi_i$
acquire vacuum expectation values (vev) $v_i$ and $u$ respectively,
breaking the $A_4$ symmetry at higher energies, and resulting in the
charged lepton mass matrix given as,
\begin{equation}
  M_{eE}M_{eE}^\dagger=\left(\begin{array}{cc}
                  (f_e v_1)^ 2 I& (f_e v_1) M_E I \\
                  (f_e v_1) M_E I & U_\omega{\rm Diag}[3(h_i^eu)^2]U_\omega^ \dagger+M_E^2 I
                  \end{array}
\right)
\end{equation}
where $f_e$ and $h_i^e$ are the Yukawa constants coupling the
standard-model fermions to the standard Higgs field and the new
scalars respectively. Here $I$ is a $3\times 3$ unity matrix and
$U_\omega$ is the magic matrix,
\begin{equation}
 U_\omega=\left(\begin{array}{ccc}
                 1 & 1 & 1 \\
                  1 & \omega & \omega^2\\
                  1 & \omega^2 & \omega
                  \end{array}
\right)
\end{equation}
with $\omega=e^{2i\pi/3}$ and we assume $v_i\ll u \ll M_E$. With such
hierarchy we have a ``universal'' see-saw scheme for generating the
standard-model charged and neutral lepton masses, that translates into
a zero-th order neutrino mixing matrix,
\begin{equation}
 U_\nu(\theta)=\left(\begin{array}{ccc}
                  \cos\theta & -\sin\theta & 0 \\
                  \sin\theta/\sqrt{2} & \cos\theta/\sqrt{2} &-1/\sqrt{2}\\
                  \sin\theta/\sqrt{2} & \cos\theta/\sqrt{2} & 1/\sqrt{2}
                  \end{array}
\right)
\end{equation}
With the discovery of nonzero $\theta_{13}$ by Daya Bay such simple
form is now excluded by experimental data, as it leads zero reactor
mixing angle due to a remnant symmetry of $A_4$. 

In this letter we focus on the generalized version of the model
proposed in~\cite{Morisi:2013qna}, by adding to it a single flavon
scalar $\xi$ that breaks this remnant symmetry present in the original
version of the model~\cite{Babu:2002dz}, and slightly changes the
charged fermion mass matrix to,
\begin{equation}
  M_{eE}M_{eE}^\dagger=\left(\begin{array}{cc}
                  (f_e v_1)^ 2 I & (f_e v_1) Y_D^\dagger \\
                  (f_e v_1) Y_D & U_\omega{\rm Diag}[3(h_i^eu)^2]U_\omega^ \dagger+Y_DY_D^ \dagger
                  \end{array}
\right)
\end{equation}
where $Y_D=M_E(I+\beta{\rm Diag}[1,\omega,\omega^ 2])$, and $\beta$ is
a small complex parameter. This equation modifies the neutrino mixing
matrix to,
\begin{equation}\label{eq:modelpred}
 U_\nu(\theta)\to K(\theta,\beta)=U_\delta^\dagger(\beta)U_\nu(\theta)
\end{equation}
where the pre-factor $U_\delta^\dagger(\beta)$ characterizes the
revamping and generates a nonzero reactor mixing angle as a result of
the breaking of the remnant $\mu-\tau$ symmetry in $A_4$. Within this
revamped scenario $|\beta|$ correlates linearly with $\theta_{13}$ and
the phase of $\beta$ induces CP violation in oscillations. Both arise
from the breaking of $\mu-\tau$ invariance.
In addition to generating these phenomenologically required
parameters, the model also predicts a correlation between the two
parameters in the lepton mixing matrix that are currently ``poorly
determined'' in neutrino oscillation studies, namely $\theta_{23}$ and
$\delta_{CP}$.

\begin{figure}[H]
    \centering
\includegraphics[scale=0.21]{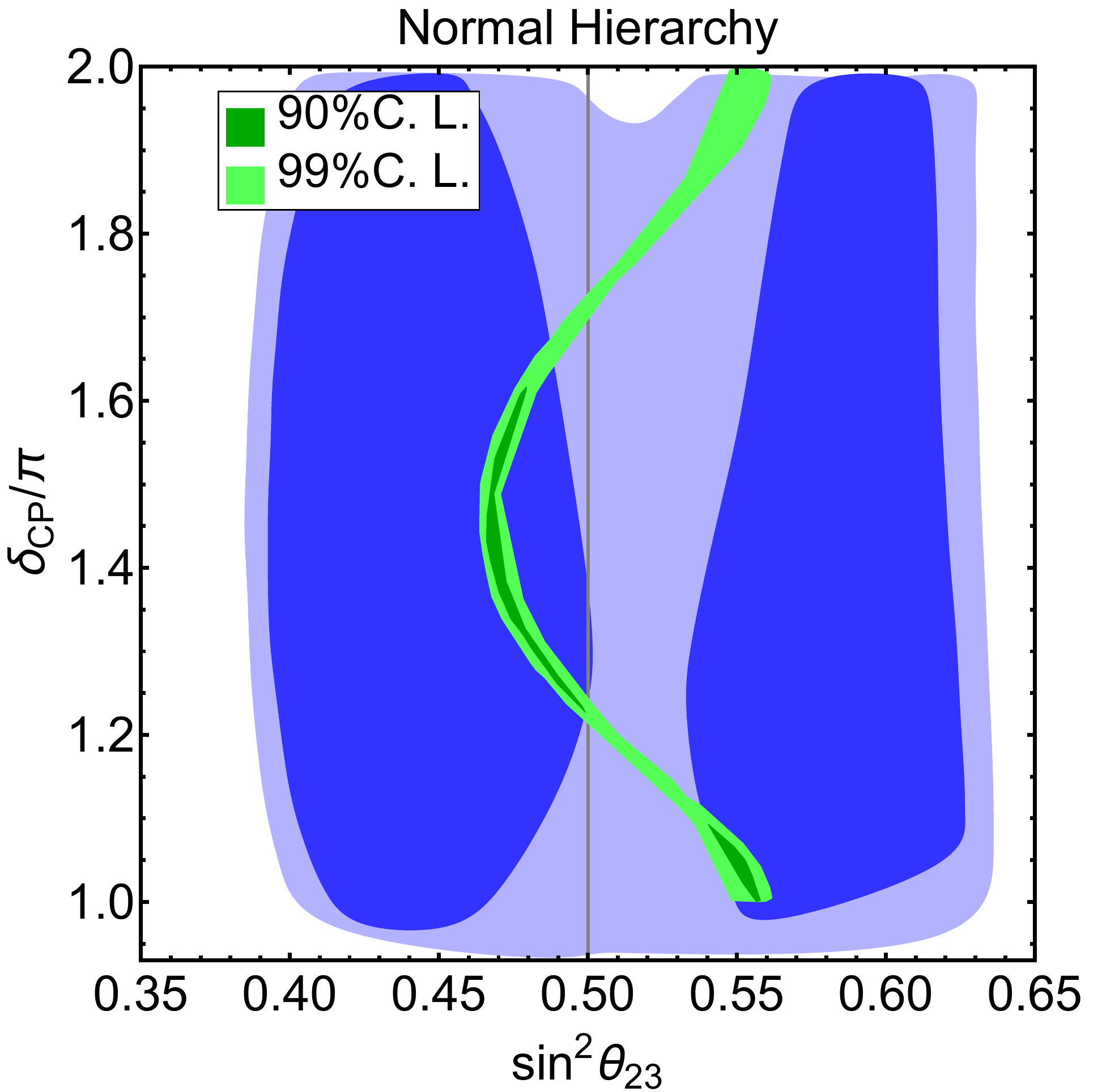}
\includegraphics[scale=0.21]{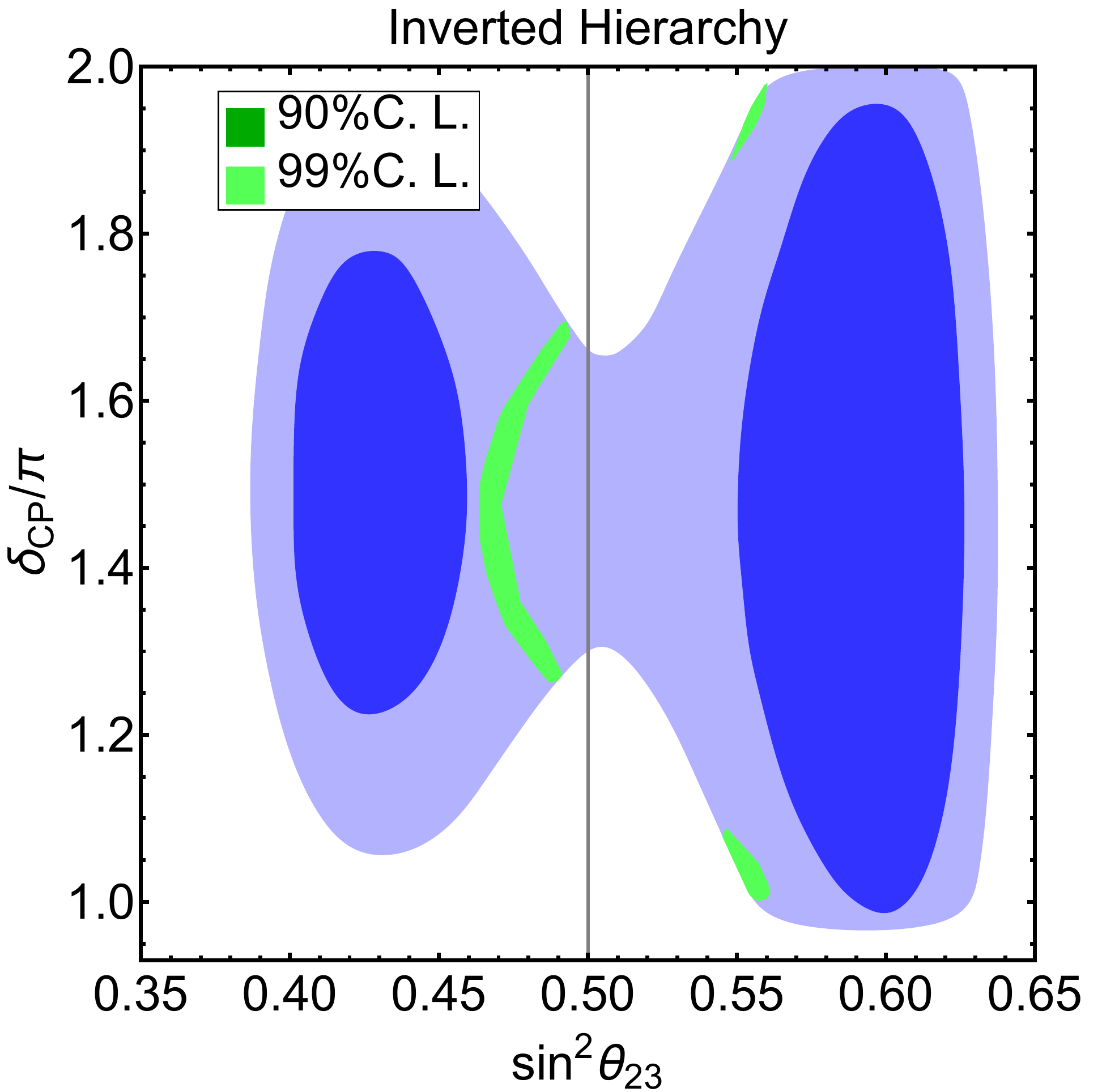}
\caption{\label{fig:current1} Regions of three-neutrino
  oscillation paramaters allowed at 90\% and 99\% of C. L. in the
  unconstrained global fit~\cite{deSalas:2017kay} (dark and light
  blue, respectively) and within the BMV scenario (dark and light green
  respectively).  The left and right panels correspond to normal and
  inverted mass ordering, respectively.}
    \end{figure}

The predicted correlation between $\theta_{23}$ and $\delta_{CP}$ can
be determined numerically by varying $|\beta|<1$,
$0\leq$Arg$[\beta],\theta\leq2\pi$, $1\leq f_ev_1\leq100$ GeV and
$10^ 4\leq M_E\leq 10^ 5$ GeV.
The results obtained are summarized in Fig.~\ref{fig:current1}, where
the dark green region indicates the predicted parameter correlation at 90\%
C.L., while the light green region is at 99\% CL.
This is a very important correlation between $\delta_{\rm CP}$
and the atmospheric angle which allows the model to be directly probed
by experiment.
It is obtained by varying the model parameters as above and by
taking only the points consistent with the current global
determination of neutrino oscillation parameters at the corresponding
confidence level.
The regions corresponding to the general unconstrained scenario given
by the latest neutrino oscillation global fit~\cite{deSalas:2017kay}
are indicated in dark and light blue, for the same confidence level.
    In contrast with the general three-neutrino oscillation picture,
    we find that, taking into account the most recent global fit of
    neutrino oscillation paramaters~\cite{deSalas:2017kay}, the
    inverted mass ordering is only allowed at the 99\% of C. L., an
    enhanced rejection than in the general unconstrained scenario.
     This is partly due to the fact that the preferred values of
     $\theta_{23}$ in the BMV case lie closer to maximality than in
     the general three-neutrino oscillation picture.

     On the other hand, the strongly preferred normal ordering case
     has two solutions, one in each octant of $\theta_{23}$.
     Of these, one notices that there is only a small region in the
     higher octant, close to a \textit{CP-conserving} value of the
     phase, $\delta_{CP}=\pi$. Although disfavored, this region is
     still allowed at 90\% of C. L., as seen by the dark green
     region.
     In contrast, the preferred solution lies in the left octant,
     close to \textit{maximal} CP violation.
     By comparing the dark green and dark blue regions one sees how the
     global analysis of the oscillation parameters within this model
     leads to an improved determination of $\theta_{23}$ and
     $\delta_{CP}$ when compared with the generic three-neutrino
     oscillation scenario.
     We now turn to the prospects of testing this model at future
     experimental setups.

\section{Numerical analysis and new experiments}
\label{simulation}

In order to determine the sensitivity of each experiment through
numerical simulation, we use the GLoBES software as described
in~\cite{Huber:2004ka,Huber:2007ji}. Unless told otherwise, the true
values of the oscillation parameters are assumed to be the best fit
values obtained in~\cite{deSalas:2017kay}, see table~\ref{tab:1}. In
accordance to recent global fit results, normal ordering has been
assumed fixed throughout the simulation.

\begin{table}[H]
 \begin{tabular}{|c|c|}
 \hline
 Parameters & \cite{deSalas:2017kay} \\ \hline \hline
$s^2_{12}$ & $0.321^{+0.18}_{-0.16}$  \\
$\theta_{12}(^{\circ})$ & $34.5^{+1.1}_{-1.0}$\\ \hline
$s^2_{13}$ & $0.0216^{+0.090}_{-0.075}$  \\
$\theta_{13}(^{\circ})$ & $8.44^{+0.18}_{-0.15}$\\ \hline
$\Delta m_{31}^2/10^{-3} ({\rm eV}^2)$ &  $2.55 \pm 0.04$ \\ \hline
$\Delta m_{21}^2/10^{-5} ({\rm eV}^2)$ & $7.56 \pm 0.19$  \\ 
\hline \hline 
$s^2_{23}$ & $0.43^{+0.20}_{-0.18}$  \\
$\theta_{23}(^{\circ})$ & $41.0 \pm 1.1$\\ \hline
$\delta_{\rm CP}/\pi$ & $1.40^{+0.31}_{-0.20}$\\
 \hline \hline 
 \end{tabular}
 \centering
 \caption{\label{tab:1}
   General three-neutrino oscillation parameters taken from the most recent global fit~\cite{deSalas:2017kay}.}
 \end{table}
 The sensitivity is calculated, at certain confidence levels, by using
 a Poissionian $\chi^2$ function~\cite{Huber:2002mx,Fogli:2002pt}
 between the true dataset $x_{i}$ and the test dataset $y_{i}$,
 \begin{eqnarray}
  \chi^2 = \min\limits_{\lbrace\xi_a,\xi_b\rbrace}\left[2\sum_{i=1}^{n}(y_i-x_i-x_i\ln\frac{y_i}{x_i})+\xi^2_a + \xi^2_b \right].
  \label{chi}
 \end{eqnarray}
 where $n$ is the total number of bins and $\xi_a$ and $\xi_b$ denote
 the pulls due to systematic errors. The test dataset is given by
 \begin{eqnarray}
  y_i(\tilde{f}, \xi_a,\xi_b) = N_i^{pre}(\tilde{f})\left[ 1+\pi^a\xi_a \right] + N_i^{b}(\tilde{f})\left[ 1+\pi^b\xi_b \right],
 \end{eqnarray}
 where $\tilde{f}$ is the set of oscillation parameters predicted by the
 model and $\pi^a$,$\pi^b$ are the systematic errors on signal and
 background respectively, assumed to be uncorrelated.

 $N_i^{pre}$ and $N_i^{b}$ represent the number of predicted
 signal events and the background events in the $i$th energy bin,
 respectively. The true or observed data assumption from an experiment
 enter in Eq.~\ref{chi} through
 \begin{eqnarray}
  x_i(f) = N_i^{obs}(f) + N_i^{b}(f),
 \end{eqnarray}
 Now the total $\chi^2$ is calculated by combining various
 relevant channels,
 \begin{eqnarray}
  \chi^2_{\rm total} = \underset{\nu_{\mu} \to \nu_e}{\chi^2} + \underset{\bar{\nu}_{\mu} \to \bar{\nu}_e}{\chi^2} +\underset{\nu_{\mu} \to \nu_{\mu}}{\chi^2} + \underset{\bar{\nu}_{\mu} \to \bar{\nu}_{\mu}}{\chi^2}~.
 \end{eqnarray}
 Finally this $\chi^2_{\rm total}$ is minimized over the free
 oscillation parameters ($\theta_{23}, \theta_{13}, \theta_{12}$, and
 $\delta_{CP}$)\footnote{Two mass squared differences have been kept
   fixed at their best fit values in Table~\ref{tab:1} since they are
   very well measured and also are not predicted by the model.}
 predicted by the model to get $\Delta \chi^{2}_{min}$.~In order to
 map out the expectations for the octant and/or CP preference we
 assume only the four ``well-measured'' oscillation parameters (upper
 rows in Table~\ref{tab:1}) plus the indication in favor of normal
 mass ordering.
   Indeed, as seen above, the inverted mass ordering is only allowed
   at the 99\% of C. L.
   In the next section we consider the case of a
   fit-independent global approach.
   We focus on two forthcoming experiments: the
   DUNE~\cite{Acciarri:2015uup} and T2HK
   experiments~\cite{Abe:2015zbg}, basing ourselves on their CDR
   report as briefly described below.\\[-.2cm]
 
 {\bf DUNE}: The proposed DUNE experiment has a baseline of 1300
   km and the far detector (FD) is placed at an on-axis location. In
   our simulation, a 40 kt liquid argon FD with 3.5 yrs. of $\nu$ run
   and 3.5 yrs. of $\bar{\nu}$ run was considered. The $\nu_{\mu}$
   beam is generated by a 80 GeV proton beam delivered at 1.07 MW with
   a POT (protons on target) of $1.47 \times 10^{21}$. The simulation
   for DUNE was done according to \cite{Acciarri:2015uup}.\\[-.2cm]
    
   {\bf T2HK}: The proposed T2HK experiment has a baseline of 295 km
   and the detector is placed at the same off-axis (0.8 degrees)
   location as in T2K. The idea is to upgrade the T2K experiment, with
   a much larger detector (560 kton fiducial mass) located in Kamioka,
   so that much larger statistics is ensured. 
   % T2HK will run for 1 year (in $\nu$ mode) + 3 years (in
   % $\bar{\nu}$ mode).
   We assume an integrated beam with power 7.5 MW $\times 10^7$
   sec. which corresponds to $1.53 \times 10^{22}$ POT. The ratio of
   the runtimes of $\nu$ and $\bar{\nu}$ mode was taken as $1: 3$.
   % The proton beam power is 7.5 MW with proton energy of 30 GeV that
   % will deliver $1.6 \times 10^{22}$ POT per year.
The simulation for T2HK was
   performed according to \cite{Abe:2015zbg}.
	
\section{DUNE and T2HK sensitivities}
\label{sec:robust-global-fit}

As seen in~\cite{deSalas:2017kay}, the atmospheric angle $\theta_{23}$
and the CP phase $\delta_{CP}$ are the two most uncertain of the
fundamental oscillation parameters.
This is in agreement with other recent global fits of neutrino
oscillations \cite{Forero:2014bxa,Esteban:2016qun,Capozzi:2017ipn}.
Theoretical scenarios, such as the BMV model, imply correlations between
them.
Thus, we now answer the very general and interesting questions:
\textit{To what extent model correlations, such as the one predicted
  by the BMV model, can be tested by experimental data? Can one
  exhibit the rejection power of future experiments independently of
  any arbitrarily given choice for the parameters $\theta_{23}$ and
  $\delta_{CP}$ eventually chosen by nature?}

\begin{figure}[t]
 \centering
\includegraphics[scale=0.4]{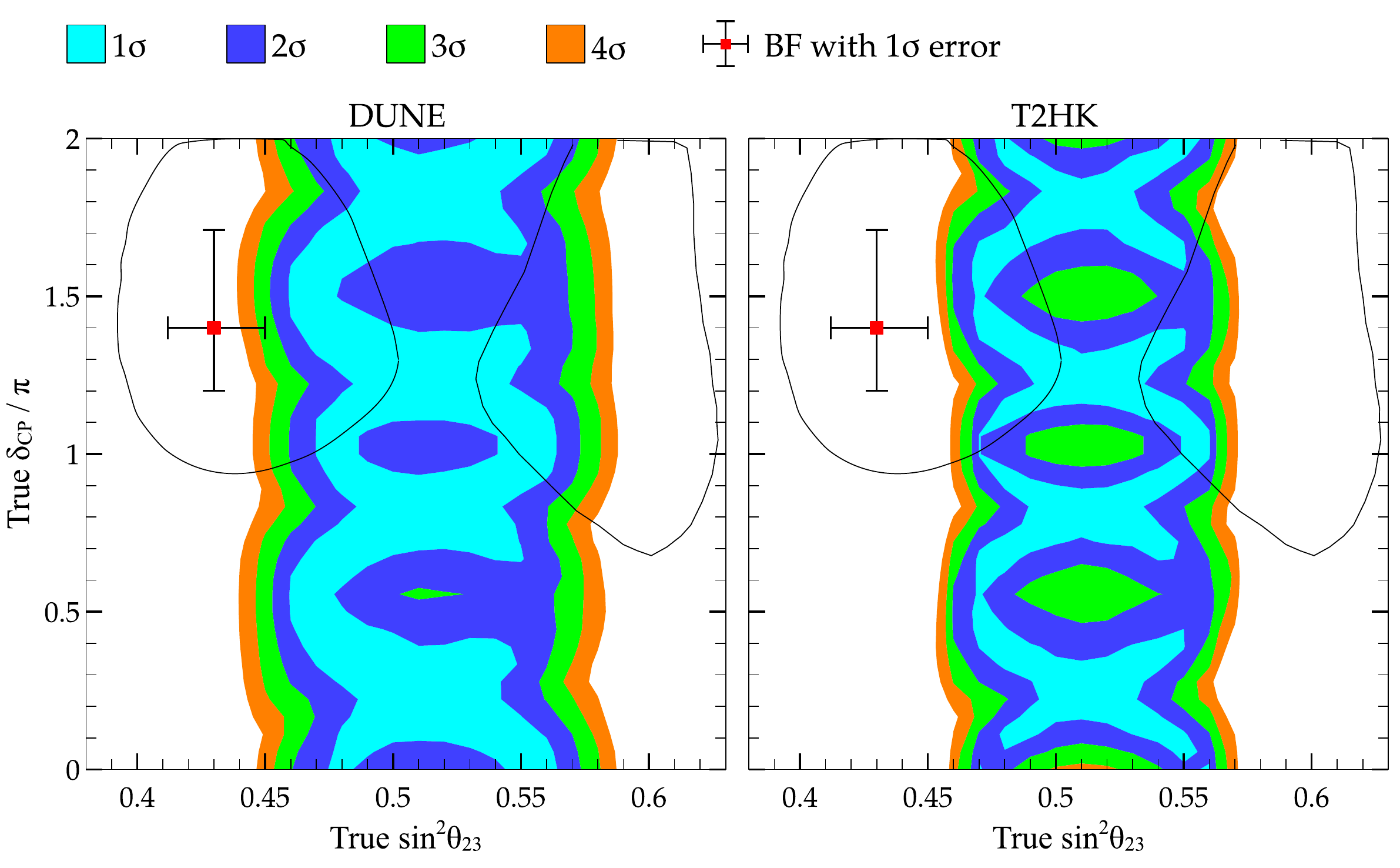}
\caption{\label{fig:scan} Expected sensitivity regions at various
  confidence levels at which DUNE (left) or T2HK (right) would test
  the revamped BMV model.  The regions within the black bordered
  contours correspond to 90\% C.L.  and the red square is the current
  best fit value~\cite{deSalas:2017kay}.  The full parameter scan of
  true values of $\sin^{2}\theta_{23}$ and $\delta_{CP}$ assumes
  normal neutrino mass ordering.  }
   \end{figure}

Performing this exercise enables us to establish robust quantitative
criteria capable of probing the model of interest, independently of
any given input from neutrino oscillation fits.
Fig.~\ref{fig:scan} answers the questions above, giving quantitative
model-testing criteria valid irrespective of any assumed global
neutrino oscillation fits.

 Our simulation procedure has been set up as follows. In order to
  calculate the oscillation parameters predicted by the model and then
  fit them to the true data set, we have marginalized over the model
  parameters within their allowed range, for each true data set.
Finally, we calculate the minimum $\Delta \chi^2$ at various
confidence levels, as shown by the different colour combinations in
Fig.~\ref{fig:scan}. The cyan, blue, green, and orange bands
correspond to the 1$\sigma$, 2$\sigma$, 3$\sigma$, and 4$\sigma$
confidence level of compatibility, at 1 degree of freedom, that is,
$\Delta\chi^2$ = 1, 4, 9, and 16 respectively. The left panel gives
the result for DUNE, while the right panel corresponds to T2HK. 
From this global-fit-independent sensitivity plot, one sees that DUNE can
  exclude, at $4\sigma$ statistical significance, the regions
  corresponding to $\sin^{2}\theta_{23} \gtrsim 0.59$ and
  $\sin^{2}\theta_{23} \lesssim 0.44$ without significant dependence
  on the value of $\delta_{CP}$ (\rm TRUE).
  On the other hand, thanks to its higher statistics, T2HK has better
  sensitivity than DUNE and consequently can exclude even larger
  regions of parameter space.
  Notice that, as indicated in both panels, the best fit point
  obtained in~\cite{deSalas:2017kay} lies outside the corresponding
  4$\sigma$ sensitivity regions at DUNE and T2HK, indicating how
  severely such parameter choice would be rejected by these
  experiments.
  We stress that these are robust model-testing criteria valid for any
  assumed global choice of neutrino oscillation parameters.
  
  \section{Summary and conclusion}
\label{sec:conclu}

Taking advantage of the latest global determination of neutrino
oscillation parameters given in~\cite{deSalas:2017kay} we have
investigated the status of the simplest revamped version of the BMV
model for neutrino oscillation, proposed in~\cite{Morisi:2013qna}, as
well as the chances of testing it further at future long-baseline
neutrino experiments.
To perform this task we have focussed on the sharp correlation between
the ``poorly determined'' oscillation parameters $\theta_{23}$ and the
phase $\delta_{CP}$ predicted in the model.
We have determined the region of these oscillation parameters allowed
within the BMV model, and compared it with what holds in the general
three-neutrino oscillation scenario.
We have found for this case a higher degree of rejection against the
higher octant of $\theta_{23}$ than in the general unconstrained case.
Through quantitative simulations of forthcoming experiments DUNE
and T2HK, according to their technical proposals, we have also
determined their potential for testing the BMV model.
We have mapped out their sensitivity regions using only the values of
the ``well-measured'' solar and atmospheric neutrino squared mass
splittings, as well as the solar and reactor angle, plus the
relatively strong preference for normal mass ordering that holds in
the BMV scenario.
We have also presented these results within a robust global approach
valid for whatever the choice of $\theta_{23}$ and $\delta_{CP}$ is
finally chosen by nature.

\section*{Acknowledgments}

Work supported by Spanish grants FPA2014-58183-P, SEV-2014-0398
(MINECO) and PROMETEOII/2014/084 (Generalitat
Valenciana). P. P. was supported by FAPESP grants 2014/05133-1,
2015/16809-9, 2014/19164-6 and FAEPEX grant N. 2391/17.

 \bibliographystyle{h-physrev}

 %\bibliography{fit-2017-v0,corfu-2017,gfit2017-refs,newrefs,merged_Valle} 

\begin{thebibliography}{}

\end{thebibliography}


\begin{thebibliography}{10}

\bibitem{deSalas:2017kay}
P.~F. de~Salas, D.~V. Forero, C.~A. Ternes, M.~Tortola, and J.~W.~F. Valle,
\newblock (2017), 1708.01186.

\bibitem{Morisi:2013qna}
S.~Morisi, D.~Forero, J.~C. Romao, and J.~W.~F. Valle,
\newblock Phys.Rev. {\bf D88}, 016003 (2013), 1305.6774.

\bibitem{1674-1137-40-10-100001}
C.~Patrignani and P.~D. Group,
\newblock Chinese Physics C {\bf 40}, 100001 (2016).

\bibitem{Boucenna:2012xb}
S.~Boucenna, S.~Morisi, M.~Tortola, and J.~W.~F. Valle,
\newblock Phys.Rev. {\bf D86}, 051301 (2012), 1206.2555.

\bibitem{Roy:2014nua}
S.~Roy, S.~Morisi, N.~N. Singh, and J.~W.~F. Valle,
\newblock Phys. Lett. {\bf B748}, 1 (2015), 1410.3658.

\bibitem{Hirsch:2012ym}
M.~Hirsch {\em et~al.},
\newblock (2012), 1201.5525.

\bibitem{Morisi:2012fg}
S.~Morisi and J.~W.~F. Valle,
\newblock Fortsch.Phys. {\bf 61}, 466 (2013), 1206.6678.

\bibitem{ishimori2012introduction}
H.~Ishimori {\em et~al.},
\newblock {\em {An Introduction to Non-Abelian Discrete Symmetries for Particle
  Physicists}}{Lecture Notes in Physics} (Springer, 2012).

\bibitem{King:2014nza}
S.~F. King, A.~Merle, S.~Morisi, Y.~Shimizu, and M.~Tanimoto,
\newblock New J.Phys. {\bf 16}, 045018 (2014), 1402.4271.

\bibitem{Chen:2015jta}
P.~Chen {\em et~al.},
\newblock JHEP {\bf 01}, 007 (2016), 1509.06683.

\bibitem{Pasquini:2016kwk}
P.~Pasquini, S.~C. ChuliÃ¡, and J.~W.~F. Valle,
\newblock Phys. Rev. {\bf D95}, 095030 (2017), 1610.05962.

\bibitem{CentellesChulia:2017koy}
S.~Centelles~ChuliÃ¡, R.~Srivastava, and J.~W.~F. Valle,
\newblock (2017), 1706.00210.

\bibitem{CarcamoHernandez:2017owh}
A.~E. CÃ¡rcamo~HernÃ¡ndez, S.~Kovalenko, J.~W.~F. Valle, and C.~A.
  Vaquera-Araujo,
\newblock JHEP {\bf 07}, 118 (2017), 1705.06320.

\bibitem{Babu:2002dz}
K.~S. Babu, E.~Ma, and J.~W.~F. Valle,
\newblock Phys. Lett. {\bf B552}, 207 (2003), hep-ph/0206292.

\bibitem{Acciarri:2015uup}
DUNE collaboration, R.~Acciarri {\em et~al.},
\newblock (2015), 1512.06148.

\bibitem{Abe:2015zbg}
Hyper-Kamiokande Proto-Collaboration, K.~Abe {\em et~al.},
\newblock PTEP {\bf 2015}, 053C02 (2015), 1502.05199.

\bibitem{Huber:2004ka}
P.~Huber, M.~Lindner, and W.~Winter,
\newblock Comput. Phys. Commun. {\bf 167}, 195 (2005), hep-ph/0407333.

\bibitem{Huber:2007ji}
P.~Huber, J.~Kopp, M.~Lindner, M.~Rolinec, and W.~Winter,
\newblock Comput.Phys.Commun. {\bf 177}, 432 (2007), hep-ph/0701187.

\bibitem{Huber:2002mx}
P.~Huber, M.~Lindner, and W.~Winter,
\newblock Nucl.Phys. {\bf B645}, 3 (2002), hep-ph/0204352.

\bibitem{Fogli:2002pt}
G.~L. Fogli, E.~Lisi, A.~Marrone, D.~Montanino, and A.~Palazzo,
\newblock Phys. Rev. {\bf D66}, 053010 (2002), hep-ph/0206162.

\bibitem{Forero:2014bxa}
D.~Forero, M.~Tortola, and J.~W.~F. Valle,
\newblock Phys.Rev. {\bf D90}, 093006 (2014), 1405.7540.

\bibitem{Esteban:2016qun}
I.~Esteban, M.~C. Gonzalez-Garcia, M.~Maltoni, I.~Martinez-Soler, and
  T.~Schwetz,
\newblock JHEP {\bf 01}, 087 (2017), 1611.01514.

\bibitem{Capozzi:2017ipn}
F.~Capozzi {\em et~al.},
\newblock Phys. Rev. {\bf D95}, 096014 (2017), 1703.04471.

\end{thebibliography}

 \end{document}